\documentclass[namedreferences]{solarphysics}

\usepackage[hyperref,optionalrh]{spr-sola-addons} 
\usepackage{graphicx}        
\usepackage{mathtools}       
\usepackage{xcolor}          
\usepackage{enumerate}

\chardef\us=`\_

\def\degr{\hbox{$^\circ$}}
\defcitealias{saito70}{S70}

\begin{document}

\begin{article}

  \begin{opening}

    \title{Radial Distributions of Coronal Electron Temperatures:
      specificities of the DYN model}

    \author[addressref={aff1,aff2},email={joseph.lemaire@uclouvain.be}]
       {\inits{J.F.}\fnm{Joseph F.}~\lnm{Lemaire}}

    \author[addressref=aff3,corref,email={katsiyannis@oma.be}]
           {\inits{A.C.}\fnm{Athanassios C.}~\lnm{Katsiyannis}
             \orcid{0000-0003-3764-0928}}

    \address[id=aff1]{Universit\'e Catholique de Louvain (UCL),
      Facult\'e des sciences, Place des Sciences, 2 bte L6.06.01, 1348
      Louvain-la-Neuve, Belgium}

    \address[id=aff2]{Royal Belgian Institute for Space Aeronomy,
      Solar-Terrestrial Centre of Excellence, Ringlaan 3, 1180,
      Belgium}

    \address[id=aff3]{Royal Observatory of Belgium, Solar-Terrestrial
      Centre of Excellence, Avenue Circulaire 3, 1180, Belgium}

    \runningauthor{J.\ Lemaire and A.\ Katsiyannis}

    \runningtitle{Coronal Electron Temperature Distribution}

    \begin{abstract}

      This paper is a follow up of the article where \citet{lemaire16}
      introduced their DYN method to calculate coronal temperature
      profiles from given radial distributions of the coronal and
      solar wind (SW) electron densities. Several such temperature
      profiles are calculated and presented corresponding to a set of
      given empirical density models derived from eclipse observations
      and in-situ measurements of the electron density and bulk
      velocity at 1 AU. The DYN temperature profiles obtained for the
      equatorial and polar regions of the corona challenge the results
      deduced since 1958 from singular hydrodynamical models of the
      SW. In these models - where the expansion velocity transits
      through a singular saddle point - the maximum coronal
      temperature is predicted to be located at the base of the
      corona, while in all DYN models the altitude of the maximum
      temperature is found at significantly higher altitudes in the
      mid-corona.  Furthermore, the maximum of the DYN-estimated
      temperatures is found at much higher altitudes over the polar
      regions and coronal holes, than over the equator.  However, at
      low altitudes, in the inner corona, the DYN temperatures are
      always smaller at high latitudes, than at low equatorial
      latitudes. This appears well in agreement with existing coronal
      hole observations.  These findings have serious implications on
      the open questions: what is the actual source of the coronal
      heating, and where is the maximum energy deposited within the
      solar corona?

    \end{abstract}

    \keywords{Corona; Inner Corona, Models; Solar Wind; Solar Wind,
      Theory; Electron Density; Electron Temperature; Coronal Heating;
      Velocity Fields, Solar Wind}
  
  \end{opening}

\section{Introduction and setting the stage for DYN model}
     \label{S-Introduction}

Measurements of White Light (WL) brightnesses and polarization (pB)
during solar eclipses have often been used in the past to infer and
calculate the electron density distribution ($\mathrm{n_e}$), the
radial distribution of coronal electrons densities, or of the
hypothetic Coronium atomic element.

According to \citet{baumbach37} pioneering analysis of coronal WL
brightnesses, $\mathrm{n_e(r)}$ can best be approximated by a sum of
terms inversely proportional powers of the r, the radial distance from
the solar center. This finding is at odds with the standard
exponential decreases, generally postulated for density profiles in
stellar and planetary atmospheres, at this epoch.

Using Baumbach's empirical formula for fitting observed coronal
densities distributions, and assuming cylindrical symmetry of the
corona around the Sun's axis of rotation, \citeauthor{saito70}
(\citeyear{saito70}, hereafter \citetalias{saito70}) constructed a
two-dimensional model for $\mathrm{n_e(r,\phi)}$ as a function of r,
and heliospheric latitude, $\mathrm\phi$.  Their empirical 2D-model is
based on a series of available eclipse observations corresponding to
epochs of minimum solar activity. \citeauthor{saito70} 's empirical
coronal electron density 2D-model became popular, and has been adopted
in many studies of the solar corona as well as in the present one,
although its range of application is restricted to r $<$ 4
$\mathrm{R_S}$ because of signal-to-noise (S/N) issues related to the
WL coronal brightnesses beyond this distance.

In order to extend \citetalias{saito70}'s density distributions up to
the Earth's orbit and beyond, \citet{lemaire16} added an extra-term
inversely proportional to the square of r. This additional power law
term fits well the solar wind distribution, whose electron density and
bulk velocity at 1 AU, will be input parameters designed hereafter by
$\mathrm{n_E}$ and $\mathrm{u_E}$, respectively.  Typical values of
these input parameters for $\mathrm{n_e(r,\phi)}$, will be chosen
within the ranges of SW observations reported by \cite{ebert09}. The
table 1 in \citet{lemaire16} contains the values of these inputs for a
set of DYN models illustrated and discussed in the present paper, as
well as in the previous one.

The analytical expression of \citetalias{saito70}'s extended density
distributions (in electrons / $\mathrm{cm^3}$) employed to determine the
temperature distributions for all DYN models is recalled here:

\begin{equation}
  \label{ne}
  \begin{aligned}
  n_e(r,\phi)= 10^8\ [3.09\ r^{-16}\ (1-0.5\sin\phi ) +
    1.58\ r^{-6}\ (1-0.95 \sin\phi ) + \\
    0.0251\ r^{-2.5}\ (1 - \sqrt{\sin\phi})] + n_E\ (215/r)^2,
  \end{aligned}
\end{equation}

\noindent where 1 AU is assumed to be 215 $\mathrm{R_S}$ and r is in
units of $\mathrm{R_S}$.

A few typical density distributions derived from this formula are
shown in figure \ref{F-density}. In order to expand the inner and
middle regions of the corona, where the greatest SW acceleration takes
place, the logarithm of h, the altitude above the photosphere -
normalised by the solar radius $\mathrm{R_S}$ - can be recommended for
the horizontal axis, and has been used in all of our graphs.

The analytical expression (\ref{ne}) happens to be a very convenient
approximation in many respects.

\begin{enumerate}[(i)]

  \item First of all, it is most convenient to determine the radial
    distributions of the coronal density gradients, and therefore that
    of H, the electron density scale height. This enables the easy
    calculation of the radial profile of the scale-height temperature,
    hereafter labelled SHM temperature, because it is determined by the
    well-known Scale-Height-Method (SHM).
 
  \item Furthermore, equation (\ref{ne}) can be used to derive an
    analytical expression for the SW bulk velocity, u(r), by
    integrating the continuity equation - the conservation of the
    particle flux - from the Earth's radial distance ($\mathrm{r_E}$)
    down to the base of the corona ($\mathrm{r_b}$), where
    $\mathrm{r_b}$ is defined hereafter to be at 1.003 $\mathrm{R_S}$.

  In the following DYN models this integration is performed along flow
  tubes whose geometrical cross-section, A(r), is an empirical
  function of r. The analytical formulas 10, 11, and 12 of
  \cite{Kopp76} are used for A(r) (for more details see section 7 and
  the appendix of \citeauthor{lemaire16}, \citeyear{lemaire16}).

  The downward integration of the hydrodynamical continuity equation
  leads to an analytical expression for u(r) defined by:
 
  \begin{equation}
    \label{E-u}
    u(r,\phi)=u_E\frac{A_E}{A(r)}\frac{n_E}{n_e(r,\phi)},
  \end{equation}  

  \noindent where $\mathrm{A_E}$ is the cross-section of the flow tube
  at 1 AU.

  To minimise the length of this paper, the mathematical formula used
  for A(r) will not be repeated here; it can be found in
  \citet{lemaire16}, equation 12. Furthermore, we will restrict
  our DYN model calculations to spherical expansions of the SW,
  i.e. $\mathrm{A_E/A(r)}$ = (215/r)$^2$.  

  We have assumed, like most other modellers of the SW that flow tubes
  of the plasma coincide with interplanetary magnetic flux tubes. This
  common assumption might be relaxed in the future, however, by
  implementing ad-hoc distributions of curl-free E-fields into the
  medium.

  \item The analytical expressions \ref{ne} and \ref{E-u} allow the
    straightforward calculation of the radial gradient of the bulk
    velocity, u(r), as well as the dimensionless function F(r),
    corresponding to the ratio of the inertial force and the
    gravitation force acting on the expanding SW plasma.

  \begin{equation}
    \label{E-Fx}
    F(r)= \frac{1}{g_SR_S}\ r^2\ u(r)\ \frac{d[u(r)]}{dr},
  \end{equation}

  \noindent where $\mathrm{g_S}$ is the gravitational acceleration at
  the solar surface (274 m s$^{-2}$).

\end{enumerate}
    
It can already be emphasized that the analytical distribution of
u(r,$\phi$) provided by equation \ref{E-u} is a continuous function of
r, and most importantly, that this function has no point of
singularity (saddle point) at the altitude where the radial expansion
of the SW becomes supersonic. This key property constitutes a major
difference between DYN models and the steady state hydrodynamical SW
models where u(r) is a singular solution of the hydrodynamical
moment/transport equations introduced by
\citet{parker58,parker63}. This issue will be discussed in greater
details in Section 6.

\section{Temperature calculation by the DYN model}
     \label{S-DYN-temperature} 

To obtain the radial distribution of the DYN temperature
\citet{lemaire16} integrated the simplest approximation of
hydrodynamic momentum transport equation, from infinity (where they
assumed that the plasma temperature is equal to zero), down to
$\mathrm{r_b}$, the base of the corona :

\begin{equation}
  \label{E-Te}
  T_e(r) = -\frac{T^*}{n_e(r)}\int_\infty^r\frac{n_e(r)}{r^2}
  \left[1+F(r)\right]dr,
\end{equation}

\noindent where $\mathrm{T^*}$ is a normalisation temperature defined
by equation 9 of \citet{lemaire16} (see also \citeauthor{alfven41},
\citeyear{alfven41}). The value of $\mathrm{T^*}$ is proportional to
the mass of the Sun, and inversely proportional to the solar
radius. It is equal to 17 MK in all of the following applications.

A non-zero additive constant temperature, $\mathrm{T_\infty}$,
corresponding to the actual electron temperature at the outer edge of
the heliosphere could have been added to the right-hand-side of
equation \ref{E-Te}. However, the addition of a constant temperature
of the order of 2000-3000 K does not change considerably the DYN
temperature profile close to the Sun, indeed $\mathrm{T_e(r)}$ is
orders of magnitude larger for $\mathrm{r_b < r < 10~R_S}$ than
$\mathrm{T_\infty}$. Therefore, it is not far from reality to set
$\mathrm{T_\infty = 0}$ as assumed in equation \ref{E-Te}.  We
verified that the DYN temperatures profiles with two widely different
values for $\mathrm{T_\infty}$ converge to the same temperatures for r
$<$ 1.1 $\mathrm{R_S}$. This remarkable convergence of the DYN
temperatures distributions at the base of the Corona, holds not only
at equatorial latitudes but also over the poles.

In a next generation of our computer code we will start the numerical
integration of equation \ref{E-Te} at $\mathrm{r_E}$, where the SW
electron temperature at 1AU, $\mathrm{T_E}$, will then be an input
parameter of the DYN model, like $\mathrm{n_E}$ and $\mathrm{u_E}$.
\citet{ebert09} is again a good source of typical $\mathrm{T_E}$
values that can then be used as a free input parameter in DYN
calculations.

Let us re-emphasized that in the DYN model the boundary conditions are
set in a large solar distance (i.e.~1 AU or infinity), and that the
continuity and momentum equations are integrated downwards. This way
DYN solutions deviate from the singular hydrodynamical solutions whose
boundary conditions are set at the bottom of the corona, and for which
the numerical integration is made upwards.

In section \ref{S-expansion} we complete the work initiated in
\citet{lemaire16} by analysing and discussing the properties of the
DYN models for other sets of the $\mathrm{u_E}$ and $\mathrm{n_E}$
input parameters. Nevertheless, it is preferable to first present and
discuss a few characteristic fits of the $\mathrm{n_e(r)}$
distribution (section \ref{S-Density}) and some properties of the DYN
temperature profiles (section \ref{S-Distributions}).

\section{Corona Electron Density Distributions inferred from 
eclipse observations.}
\label{S-Density}

  \begin{figure}
    \centerline{\includegraphics[width=\textwidth,clip]
      {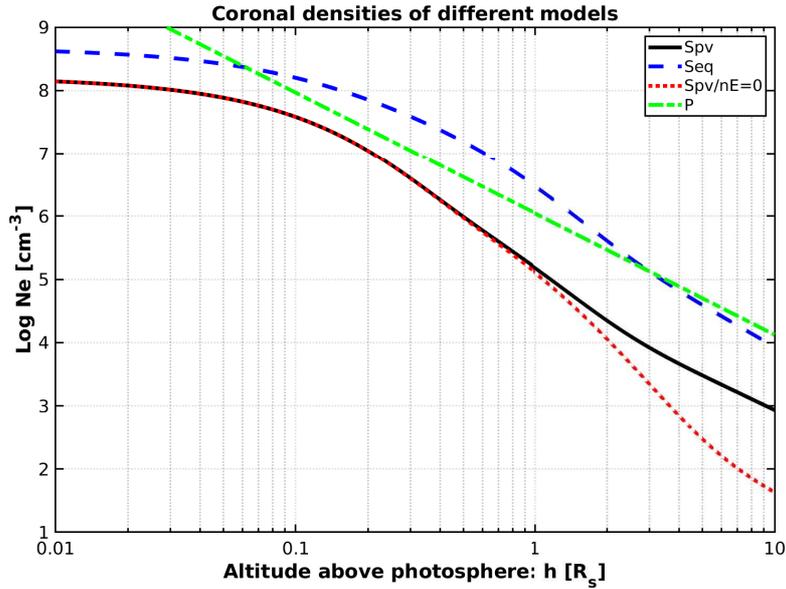}}
    \caption{Expanded coronal electron density distributions for the
      equatorial (blue-dashed curve, Seq), and polar regions described
      by \citet{lemaire16}  (black curve, Spv), taken from
      \citetalias{saito70} (red dotted curve, Spv/nE=0), and from
      \cite{pottasch60} (green-dashed-dotted curve, P). Pottasch's
      density distribution was also included in figure 6.5 of
      \cite{parker63}.}
    \label{F-density}
  \end{figure}

The electron density distribution, $\mathrm{n_e(r)}$, implemented by
different authors from eclipse observations is displayed in figure
\ref{F-density}. This graph is similar to the figure 1 of
\citet{lemaire16} . It is included here for completion and easier
access.

The black curve (Spv) corresponds to Saito's polar density
distribution which has been extended to large distances by adding the
contribution of the Solar Wind density.  The red-dotted curve
(Spv/nE=0) is the same distribution but without the last term of
equation (\ref{ne}), i.e.\ without the contribution of the solar wind
density. The DYN temperature determined for this (red) density
distribution corresponds the HST temperature profile for which which
u(r)= 0. Indeed, according to equation (\ref{E-u}), u(r) is
proportional to $\mathrm{n_E}$, thus u(r)=0 when $\mathrm{n_E}$=0.
This (red) density profile can thus be viewed as a radial density
distribution wherein the class of escaping SW particles would be
missing in the exospheric coronal models of
\cite{lemaire71,lemaire73}.

In collisionless/kinetic models, it is exclusively the class of
escaping electrons that contributes to the net outward flux of
evaporating coronal electrons. The classes of ballistic and trapped
particles don't contribute to this net SW flux of particles. Indeed
these latter electrons do not have sufficiently large energy to escape
out Lemaire-Scherer's electrostatic potential well.  Nevertheless,
they play a major role by contributing their electric charges to the
total negative charge density of the coronal and SW plasma. Note also
that these collisionless ballistic and trapped particles do not
contribute either to the net outward flux of kinetic energy that is
carried out of the corona into interplanetary space by the SW.  Thus
these low energy (sub-thermal and thermal) electrons contribute
exclusively to the total negative charge density which must balance
the positive charge density of the ions, in order to keep the plasma
locally quasi-neutral. In contrast, in fluid or hydrodynamical
representations of the coronal and SW plasma no such discrimination
between sub-thermal and supra-thermal electrons, or between escaping,
ballistic and trapped electrons is made explicitly. All classes of
electrons are assumed to take part to the net outward fluxes of SW
particles and of the SW kinetic energy. This constitutes a fundamental
distinction between both types of plasma representations. These key
differences are discussed in greater details in the review article by
\citet{echim11}.

The blue-dashed curve (Seq) corresponds to \citetalias{saito70}'s
extended equatorial density model (i.e. for ~$\mathrm\phi = 0$). Note
that the latter equatorial density (Seq) is significantly larger than
the the polar density distribution (Spv) in the inner and middle
corona.

The green curve (P) corresponds to a best fit of an equatorial
electron density distribution during solar minimum determined by
\citet{pottasch60}. It was derived from WL brightness and polarization
measurements during the solar eclipse of 1952, under the assumption
that the corona would be in hydrostatic equilibrium.

Although not stressed any further here, we noted that the density
profiles associated with the critical solutions of the SW hydrodynamic
momentum/ transport equations have significantly smaller density
gradients (i.e. significantly larger density scale-heights) in the
inner corona than the empirical models shown in the figure
\ref{F-density}, which were derived from of eclipse observations. To
our knowledge this misfit has generally been overlooked, except in
figure 1 of \citet{scarf65}.

\section{Properties of the DYN temperature profiles}
\label{S-Distributions}

  \begin{figure}
    \centerline{\includegraphics[width=\textwidth,clip]
      {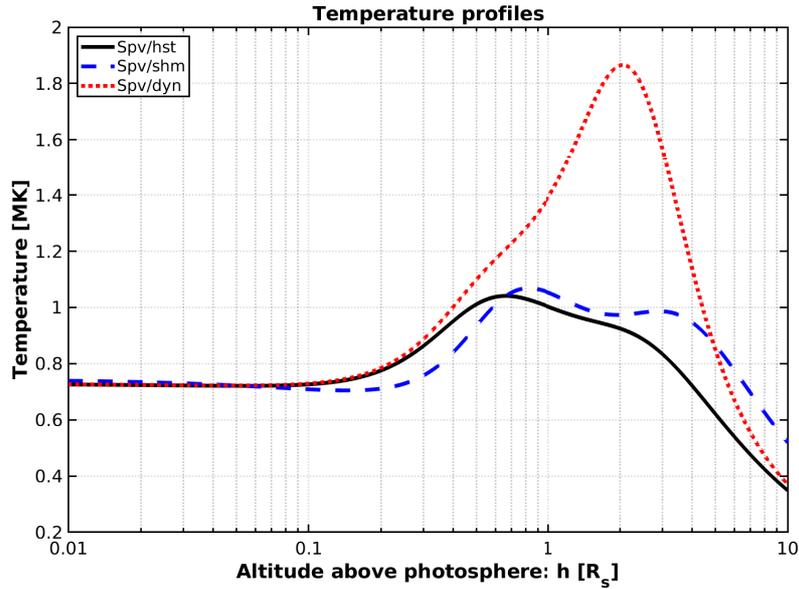}}
    \caption{The electron temperature profiles over the polar regions
      (Spv) calculated by using the three different methods: the scale
      height (SHM), the hydrostatic (HST), and the hydrodynamical
      (DYN). All three curves are calculated with the same polar
      electron density profile (Spv; the black-curve in figure 1) for
      which $\mathrm{n_E}$= 2.2 cm$^{-3}$ at 1 AU. The red dotted
      curve (Spv/DYN) is obtained by assuming that the SW velocity at
      1 AU is equal to $\mathrm{u_E}$ = 329 km/s (which is an average
      value for slow SW flows), while the solid black-curve (Spv/HST)
      is obtained for $\mathrm{u_E}$ = 0; the DYN model coincides then
      with the hydrostatic HST model).}
    \label{F-temperature}
  \end{figure}

The three electron temperatures profiles shown in figure
\ref{F-temperature} are obtained for the same polar density profile,
Spv, by using the three different methods of calculation (SHM, HST,
and DYN method) recalled above. This polar density profile is
illustrated by the black curve in figure \ref{F-density}. It
corresponds to \citetalias{saito70}'s expended polar density
distribution (~$\mathrm\phi = 90\degr$) with $\mathrm{n_E}$ = 2.22
electrons/cc at 1 AU. A similar trio of temperature profiles were
shown in figure 3 of \citet{lemaire16}, obtained for the equatorial
density distribution, Seq, corresponding to \citetalias{saito70}'s
extended equatorial density distribution with $\mathrm{n_E}$ = 5.75
electrons/cm$^3$ at 1 AU (i.e. the blue dashed curve in figure
\ref{F-density}).
 
In the DYN models shown in figure \ref{F-temperature} and in all the
following it is assumed that the temperature of the coronal protons is
the same as that of the electrons ($\mathrm{T_p/T_e = \tau_p =
  1}$). Furthermore, it is assumed that the concentration of heavier
ions is equal to zero ($\mathrm{n_{He^{++}}/n_{H^+} = \alpha = 0}$).
However, these questionable simplifications can easily be relaxed. In
the current MATLAB\circledR\ code developed by \citet{lemaire16},
the value of $\alpha$ and $\mathrm{\tau_p}$ can be given different
constant values which are independent of r. Results for such more
evolved DYN models have been reported in Table 2 of \citet{lemaire16},
and will not be repeated here.

Comparing both figures (figure \ref{F-temperature} and figure 3 of
\citeauthor{lemaire16}, \citeyear{lemaire16}) it can be seen that:

\begin{enumerate}[(i)]

  \item The maximum value of the SHM temperature distribution is
    always situated at a somewhat higher altitudes than the maximum
    value of the HST temperature. Indeed in the SHM method of
    calculation of $\mathrm{T_e(r)}$, the effect of the temperature
    gradient, $\mathrm{dT_e(r)/dr}$, is ignored (see
    \citeauthor{lemaire16}, \citeyear{lemaire16}), while it is
    properly taken into account in the HST method first developed by
    \citet{alfven41}.

  \item Both the SHM and the HST methods give maximum values,
    $\mathrm{T_{e,max}}$, that are nearly equal to each other (circa 1
    MK, over the poles, and slight larger than 1.2 MK, over the
    equator).

  \item the maximum of the DYN temperature is much larger than the
    maximum of the HST temperature over the poles (see figure
    \ref{F-temperature}), while over the equatorial region, these two
    temperature maxima are almost identical (see figure 3 of
    \citet{lemaire16}); this is, of course, a consequence of the much
    larger coronal density over the equator.

  \item The DYN and HST methods give almost identical temperatures
    profiles at low altitudes, for $\mathrm{h < 0.1~R_S}$. This result
    is clearly foreseeable because at these lowest altitudes in the
    inner corona the coronal plasma is almost in hydrostatic
    equilibrium i.e. u(r) $\approx$ 0 both over the polar
    and equatorial regions.

  \item In the inner corona the temperature gradients are positive :
    $\mathrm{dT_e(r)/dr}~>$ 0, but they tend to become smaller and
    smaller when r decreases to $\mathrm{r_b}$. This is a basic
    property satisfied both over the equator and the poles by all DYN
    solutions.  This trend is, however, at odds with the temperature
    gradients predicted by the usual singular solutions of the
    hydrodynamical transport equations. Indeed in the latter critical
    hydrodynamical models the coronal temperature is in general a
    decreasing function of r, even at the base of the corona.

\end{enumerate}

As a consequence of the much lower densities over the poles than over
the equator, the radial distributions of the DYN and HST temperatures
begin to depart from each other at much lower altitudes over the
poles, than over the equatorial region. This occurs at h $\gtrapprox$
0.2 $\mathrm{R_S}$ at high latitudes, while only at h $\gtrapprox$ 1
$\mathrm{R_S}$ (i.e.\ r $\gtrapprox$ 2 $\mathrm{R_S}$) over the lower
equatorial latitudes.

any of the trends of the DYN temperatures outlined above and
illustrated in figure \ref{F-temperature}, as well as in figure 3 of
\citet{lemaire16}, are consistent with the observed properties of the
solar corona temperatures reported in reviews, such as
\citet{echim11}. 

The ongoing solar missions carry new kinds of instruments, such as the
Wide-field Imager for Solar Probe (WISPR, \citeauthor{vourlidas16},
\citeyear{vourlidas16}) and Metis \citep{antonucci19}, capable of
observing the $\mathrm{n_e(r)}$ distribution far more accurately and
much more regularly than previously.  The first results from WISPR are
already published by \citet{howard19b} and their figure 1 is in-line
with S70's density profiling. More results from those instruments are
greatly anticipated by the authors as valuable input to the DYN
models.

\section{The effects of the input parameters $\mathrm{n_E}$ and
  $\mathrm{u_E}$ on the DYN temperature distribution}
\label{S-expansion}

  \begin{figure}
    \centerline{\includegraphics[width=\textwidth,clip]
      {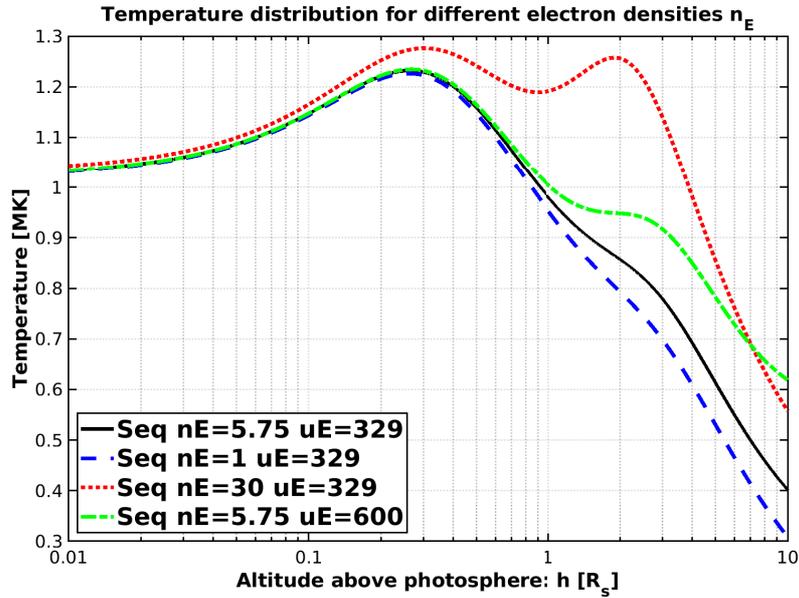}}
    \caption{Equatorial distributions of the DYN-temperature obtained
      for the following sets of the SW bulk velocity and electron
      density at 1 AU: $\mathrm{n_E}$= 1 $\mathrm{e^-/cm^3}$,
      $\mathrm{u_E}$ = 329 km/s (blue-dashed line); $\mathrm{n_E}$
      =5.75 $\mathrm{e^-/cm^3}$, $\mathrm{u_E}$ = 329 km/s
      (black solid line); $\mathrm{n_E}$ = 30 $\mathrm{e^-/cm^3}$,
      $\mathrm{u_E}$ = 329 km/s (red dotted line); and $\mathrm{n_E}$
      = 5.75 $\mathrm{e^-/cm^3}$, $\mathrm{u_E}$ = 600 km/s
      (dashed green line).}
    \label{F-nE}
  \end{figure}

The figure 3 displays the DYN temperature profiles based on
four different equatorial density profiles, $\mathrm{n_e(r,\phi)}$,
corresponding to Saito's extended density models for
$\mathrm\phi=~90\degr$, $\mathrm{n_E}$ = 1.0; 5.75; 30.0
$\mathrm{e/cm^3}$, and $\mathrm{u_E}$ = 329; 600 km/s.

For the black curve in figure \ref{F-nE}, $\mathrm{u_E}$ = 329 km/s
and $\mathrm{n_E}$ = 5.75 $\mathrm{e/cm^3}$. These input parameters
are respectively the SW bulk velocity and number density at 1 AU of
the average slow SW flow (reported by \citeauthor{ebert09},
\citeyear{ebert09}). Two other curves (blue dashed and red dotted)
show the DYN temperatures respectively for smaller and larger values
of $\mathrm{n_E}$.  It can be seen that when the solar wind density at
1 AU is reduced (from $\mathrm{n_E}$ = 5.75 to 1 $\mathrm{e/cm^3}$),
the temperature profile is slightly reduced and tends to the HST
temperature distribution. This can be see by comparing this curve to
the Seq/HST curve in figure 3 of \citet{lemaire16}; indeed, the latter
was calculated by using the HST method introduced by
\citet{alfven41}. It comes as no surprise that the two are identical,
since for a negligible amount of SW at 1 AU the DYN model becomes
equivalent to the HST model.

However, when the SW density at 1AU is arbitrarily enhanced
($\mathrm{n_E}$ = 30 $\mathrm{e/cm^3}$ or more) it can be seen that
a higher value of the maximum temperature is obtained in the
mid-corona as expected to boost the coronal plasma to
a bulk speed of $\mathrm{u_E}$ = 329 km/s or more, at 1 AU.

The green dotted curve in figure \ref{F-nE} has a bump at h = 2
$\mathrm{R_S}$. This leads to the evidence that an enhanced maximum
temperature, and thus an enhanced coronal heating rate, is required in
the mid-corona in order to boost the SW speed at 1 AU up from 329 km/s
(black curve corresponding to a slow SW flow) to 600 km/s (green
curve; corresponding a fast wind speed).

The remarkable convergence at low altitudes of all four curves shown
in figure \ref{F-nE} tells us that the coronal temperature in the
inner corona is almost unaffected by $\mathrm{n_E}$ and
$\mathrm{u_E}$, the SW density and speed at 1 AU.  It is basically the
maximum DYN temperature within the mid-corona that determines the SW
at 1 AU and in the distant interplanetary medium. In other words to
enhance the SW expansion velocities or/and to enhance the plasma
densities in the interplanetary medium, increased heating is not
required at the base of the corona, but higher up in the corona at a
radial distance of 3-4 $\mathrm{R_S}$. This is a most important new
finding grounded on our DYN model calculations.

\section{The differences between DYN models and Parker's hydrodynamical models}
\label{S-boundary}

The fundamental limitations of the hydrostatic coronal models are well
understood. They were first pointed out in the papers by
\citet{parker58,parker63} The hydrodynamical plasma transport
equations he introduced were integrated upwards from a low altitude
reference level, $\mathrm{r_0}$, up to infinity.  Very precisely
chosen boundary condition, $\mathrm{u_0}$, had to be chosen at
$\mathrm{r_0}$ to obtain a continuous solution for u(r) crossing a
saddle point at the altitude where the SW expansion velocity becomes
supersonic. Any other slightly different boundary conditions would
produce diverging steady state solutions. In the 60's this critical
solution for the SW flow velocity had been compared to the similar
hydrodynamical solution describing the supersonic flow velocity in a
de Laval nozzle.

On the contrary, the DYN distributions of u(r) are continuous but
non-singular solutions of the continuity and momentum equations
describing the SW expansion. Indeed they are not characterized by a
saddle point.  In the DYN models these hydrodynamical transport
equations are integrated downwards from 1 AU, where appropriate
boundary conditions are taken as free input parameters. As indicated
above this has the remarkable advantage to generate wide ranges of
continuous solutions for the SW expansion velocity, and for the
electron temperature distributions. Furthermore, all the latter
DYN temperature profiles happen to converge at the base of the corona
to HST temperature which corresponds to the hydrostatic model,
whatever the values of $\mathrm{n_E}$ and $\mathrm{u_E}$, may have
been assumed at 1 AU.

The singular solutions crossing a saddle point, lead to coronal
temperatures that maximize at the base of the corona, having negative
temperature gradients in the inner corona. But this is in contrast to
the DYN temperature profiles which predict always that the values of
$\mathrm{dT_e(r)/dr}$ are positive and small, both over the coronal
poles (see figure \ref{F-temperature}), and over the equator (see
figure \ref{F-nE}).

Note however, these nearly uniform values of $\mathrm{T_e(r)}$, in the
inner corona, are larger in equatorial region (1.0-1.3 MK), than at
high latitudes, over the poles and in coronal holes (0.7-0.8 MK).
This remarkable difference of temperatures between the equatorial and
polar in the inner corona, is well supported by SKYLAB and other EUV
and X-ray observations.

Albeit in the inner corona ($\mathrm{h < 0.1 R_S}$) the nearly
isothermal polar temperatures are much lower than the equatorial ones,
the DYN models predict that the reverse is true at higher altitudes:
i.e. in the mid-corona for h = 2-3 $\mathrm{R_S}$, when realistic
values are adopted for $\mathrm{n_E}$ and $\mathrm{u_E}$.

From the results presented above, it can be seen that at such higher
altitudes the maximum of the DYN temperatures is then much larger over
the poles or in coronal holes ($\mathrm{T_{e,max}(r)}$ = 1.8-2.5 MK),
than over the equatorial regions ($\mathrm{T_{e,max}(r) < 1.3}$ MK).
This result implies, thus, that much larger electron temperatures are
indeed needed at mid-altitudes in coronal holes to accelerate SW
streams to high speeds (600 km/s or more), than is needed over the
equatorial regions from where the slower SW streams (330 km/s or so)
are suspected to originate.

This conclusion is fully consistent with Parker's expectation in the
60's and \cite{lemaire71,lemaire73}'s one in the 70's that larger
coronal temperatures would necessarily be required in the corona, to
boost the coronal plasma to larger speeds in interplanetary
space. This leads us to infer that larger energy deposition rates are
needed in the mid-corona, but not necessarily at lower altitudes
inside the inner-corona.

Albeit in the present paper we do not address the pending issue of
possible coronal heating mechanisms able to account for the DYN
temperature profiles displayed in figures \ref{F-temperature} and
\ref{F-nE}, we wish to insist that these profiles have their maximum
in the mid-corona, not in the transition region at
$\mathrm{h_{tr}\approx 0.003~R_S}$.

\section{Conclusions}
\label{S-Conclusions}

The calculated DYN temperature distributions have been compared with
those determined by using older methods of calculation, especially,
the SHM commonly used by assuming a corona in hydrostatic equilibrium
and in isothermal equilibrium. It has been shown that the latter
untenable assumptions are leading to coronal electron temperature
distributions that are quite different from those obtained by the DYN
method introduced by \citeauthor{lemaire16}.

The DYN model is a straightforward extension of the hydrostatic model
developed decades ago by \citet{alfven41}. Indeed, it this more
general model takes into account the radial expansion of the coronal
plasma, without any transverse motion of plasma across magnetic field
lines. Indeed, here it has been assumed that the coronal plasma is
flowing up in open flow tubes that coincide with magnetic flux tubes
whose geometry and cross-section, A(r), are the same as those adopted
by \cite{Kopp76}. Here A(r) is an ad-hoc analytic input function
associate to the DYN model.

The radial electron density distribution, $\mathrm{n_e(r)}$, is an
additional input function required to create a DYN model. It can be
derived from WL eclipse observations, or from an empirical model like
that of \citeauthor{saito70}.  Unfortunately, for A(r) there does not
yet exist such a ``steering oar'' to guide our ``educated guesses''.

After having pointed out the major differences between the temperature
profiles obtained by the new DYN method in comparison to the SHM and
HST models, we have shown that in all cases the calculated coronal
temperatures have a maximum value in the mid-corona, and never at the
base of the corona, as often implied in publications.

This important finding has lead us to the conjecture that the source
of the coronal heating is not at the base of the corona, but higher up
in the mid-corona, where the DYN temperatures distributions have a
well defined maximum at all heliospheric latitudes. Note that even the
HST temperature and SHM temperature profiles have temperature maxima
well above the base of the corona.

These theoretical results put in question the common hypothesis that
the corona is heated exclusively from below. Indeed, although widely
spread, this believe is only a hypothesis based on the reasonable
expectation that heating takes place where the energy density is at
its highest. However, to the best of the authors knowledge, there are
no evidence to rule out the possibility that a significant amount of
heating takes place higher up.

Conversely, the DYN model cannot rule out the existence of the
commonly-suggested heating mechanisms (such as those based on
reconnection or magneto-hydrodynamical dumping). As it can be seen in
figures \ref{F-temperature} and \ref{F-nE} and also in
\citet{lemaire16}, the temperature calculated at $\mathrm{r_b}$ is
much higher than the observed chromospheric temperatures.

Obviously this cannot be considered as the end of the SW modelling
venture but it is nevertheless a basic new step ahead. Much more
elaborate and difficult work stays to model the kinetic pressure
anisotropies of the electrons and ionic populations as well as their
mutual collisional interactions, the most important and interesting
challenge remaining, of course, the determination of the coronal heat
deposition rate versus heliospheric distances and latitudes.

\begin{acks}

We acknowledge the logistic support of BELSPO, the Belgian Space
Research Office, as well as the help of the IT teams of BIRA and ROB.
JFL wishes also to thanks Viviane Pierrard (BIRA-IASB), Marius Echim
(BIRA-IASB), Koen Stegen (ROB), and the early assistance of Cl\'ement
Botquin, IT student hired during the Summer 2010.  We acknowledge
Serge Koutchmy for his interest in our work, and for pointing out in
page 3664 of \citeauthor{lemaire16} a mistake in the given numerical
value of H (the density scale height at the base of the Corona when
the temperature is assumed to be equal to 1 MK). ACK acknowledges
funding from the Solar-Terrestrial Centre of Excellence (STCE), a
collaborative framework funded by the Belgian Science Policy Office
(BELSPO). Some work for this paper was also done in the framework of
the SOL3CAM project (Contract: BR/154/PI/SOL3CAM), funded by BELSPO.

The authors are not aware of any conflicts of interest. The authors
have no relevant financial or non-financial interests to disclose. All
work done for this article was funded by the Royal Belgian Institute
for Space Aeronomy (BIRA-IASB), and the Royal Observatory of Belgium
(ROB). Both institutes are funded by the BELgian Science Policy Office
(BELSPO).

\end{acks}

\bibliographystyle{spr-mp-sola}
\bibliography{katsiyannis}  

\IfFileExists{\jobname.bbl}{} {\typeout{}
\typeout{****************************************************}
\typeout{****************************************************}
\typeout{** Please run "bibtex \jobname" to obtain} \typeout{**
the bibliography and then re-run LaTeX} \typeout{** twice to fix
the references !}
\typeout{****************************************************}
\typeout{****************************************************}
\typeout{}}

\end{article} 

\end{document}